\documentclass[12pt]{emulateapj}

\usepackage[breaklinks,colorlinks,urlcolor=blue,citecolor=blue,linkcolor=blue]{hyperref}
\usepackage{apjfonts}
\usepackage{amssymb,amsmath}
\usepackage{natbib}
\usepackage{graphicx}
\newcommand{\teff}{${T}_{\mathrm{eff}}$}
\newcommand{\logg}{$\log{g}$}
\newcommand{\msun}{$M_{\odot}$}
\newcommand{\mstar}{${M}_{\star}$}

\newcommand{\kms}{km~s$^{-1}$}

\newcommand{\muhz}{$\mu$Hz}

\newcommand{\tar}{SDSSJ0837+1856}

\shorttitle{1.13-hr rotation in a 0.87 \msun\ white dwarf}
\shortauthors{Hermes et al.}

\begin{document}

\title{EVIDENCE FROM {\em K2} FOR RAPID ROTATION IN THE DESCENDANT OF AN INTERMEDIATE-MASS STAR}
\author{J.~J.~Hermes\altaffilmark{1,2}, Steven~D.~Kawaler\altaffilmark{3}, A.~D.~Romero\altaffilmark{4}, S.~O.~Kepler\altaffilmark{4}, P.-E.~Tremblay\altaffilmark{5}, Keaton~J.~Bell\altaffilmark{6}, B.~H.~Dunlap\altaffilmark{1}, M.~H.~Montgomery\altaffilmark{6}, B.~T.~G\"{a}nsicke\altaffilmark{5}, J.~C.~Clemens\altaffilmark{1}, E.~Dennihy\altaffilmark{1}, and S.~Redfield\altaffilmark{8} }

\altaffiltext{1}{Department of Physics and Astronomy, University of North Carolina, Chapel Hill, NC 27599, USA}
\altaffiltext{2}{Hubble Fellow}
\altaffiltext{3}{Department of Physics and Astronomy, Iowa State University, Ames, IA 50011, USA}
\altaffiltext{4}{Instituto de F\'{\i}sica, Universidade Federal do Rio Grande do Sul, Porto Alegre, RS, Brazil}
\altaffiltext{5}{Department of Physics, University of Warwick, Coventry CV4~7AL, UK}
\altaffiltext{6}{Department of Astronomy, University of Texas at Austin, Austin, TX 78712, USA}
\altaffiltext{7}{Wesleyan University Astronomy Department, Van Vleck Observatory, 96 Foss Hill Drive, Middletown, CT 06459, USA}

\email{jjhermes@unc.edu}

\begin{abstract}

Using patterns in the oscillation frequencies of a white dwarf observed by {\em K2}, we have measured the fastest rotation rate ($1.13\pm0.02$\,hr) of any isolated pulsating white dwarf known to date. Balmer-line fits to follow-up spectroscopy from the SOAR telescope show that the star (SDSSJ0837+1856, EPIC\,211914185) is a $13{,}590\pm340$\,K, $0.87\pm0.03$\,\msun\ white dwarf. This is the highest mass measured for any pulsating white dwarf with known rotation, suggesting a possible link between high mass and fast rotation. If it is the product of single-star evolution, its progenitor was a roughly 4.0\,\msun\ main-sequence B star; we know very little about the angular momentum evolution of such intermediate-mass stars. We explore the possibility that this rapidly rotating white dwarf is the byproduct of a binary merger, which we conclude is unlikely given the pulsation periods observed.
\end{abstract}


\section{Introduction}
\label{sec:intro}

Thanks to long-baseline monitoring enabled by space missions like {\em CoRoT} and {\em Kepler}, we now have deep insight into the angular momentum evolution of low-mass stars \citep{2015AN....336..477A}. Asteroseismology enables measuring core and surface rotation rates for numerous $1-3$\,\msun\ stars along the main sequence (e.g., \citealt{2017MNRAS.465.2294O,2016A&A...593A.120V} and references therein) and along their first ascent up the red giant branch (e.g., \citealt{2012A&A...548A..10M}), as well as in core-helium-burning, secondary clump giants \citep{2015A&A...580A..96D}.

As powerful as the {\em Kepler} seismology has been, it has so far determined internal rotation rates for just a few intermediate-mass stars ($3$~$<$~$M$~$<$~$8$\,\msun) on the main sequence (e.g., \citealt{2017A&A...598A..74P} and references therein). Therefore we have few constraints on the past or future evolution of angular momentum in Cepheids; rotation can have a significant evolutionary impact on these standard candles \citep{2014A&A...564A.100A}.

As with all low-mass stars, intermediate-mass stars below roughly 8\,\msun\ will end their lives as white dwarfs. We can therefore constrain the final stages of angular momentum evolution of intermediate-mass stars by observing white dwarfs. The majority of field white dwarfs have an overall mass narrowly clustered around 0.62\,\msun, as determined by fits to the pressure-broadened Balmer lines of hydrogen-atmosphere (DA) white dwarfs \citep{2016MNRAS.461.2100T}. Initial-final mass relations calibrated using white dwarfs in clusters suggest that 0.62\,\msun\ white dwarfs evolved from roughly 2.2\,\msun\ main-sequence progenitors (e.g., \citealt{2009ApJ...693..355W}).

Currently known white dwarf rotation rates lead to expected rotational broadening well below currently measured upper limits, usually $v\sin{i}<10$\,\kms\ \citep{2005A&A...444..565B}. Thus, our best insights into the rotation rates of white dwarfs evolving without binary influence come from asteroseismology. To date, rotation of roughly 20 white dwarfs has been measured from their pulsations, with rotation periods spanning $0.4-2.2$\,d \citep{2015ASPC..493...65K}. All but two have masses less than $0.73$\,\msun, suggesting they generally represent the endpoints of $<$$3.0$\,\msun\ progenitors.

Thanks to its tour of many new fields along the ecliptic, the second-life of the {\em Kepler} space telescope, {\em K2}, is rapidly increasing the number of white dwarfs with nearly uninterrupted, multi-month light curves suitable to measure interior rotation rates in these stellar remnants (e.g., \citealt{2017ApJ...835..277H}). We present here the discovery of pulsations in a massive white dwarf --- likely the descendent of a roughly 4.0\,\msun\ main-sequence star --- which we find to be the most rapidly rotating pulsating white dwarf known to date.

\begin{figure*}[t]
\centering{\includegraphics[width=0.995\textwidth]{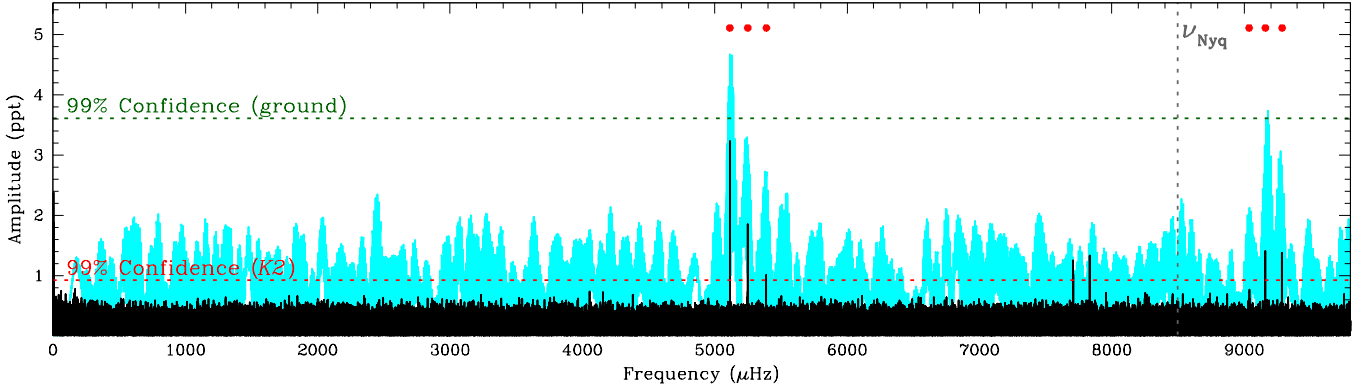}}
\caption{The Fourier transform of the {\em K2} data of \tar\ (shown in black) shows six frequencies of stellar variability, marked with red dots. A Fourier transform of follow-up ground-based photometry from the 2.1-m Otto Struve telescope at McDonald Observatory and the 4.1-m SOAR telescope, shown in cyan, resolve the Nyquist ambugity caused by the 58.8\,s {\em K2} short-cadence exposures. The pulsation modes are shown in more detail in Figure~\ref{fig:triplets}. \label{fig:ft}}
\end{figure*}

\section{Pulsation Periods from {\em K2}}
\label{sec:photobs}

We targeted the star SDSSJ083702.16+185613.4 (hereafter \tar, EPIC\,211914185) during {\em K2} Campaign~5 as part of a search for transits around white dwarfs using the shortest-cadence observations possible (program GO5073). \tar\ was not observed as part of our GO program to search for pulsations in white dwarfs, as we believed it too faint ($K_p=18.9$ mag). It was selected as a candidate white dwarf from the photometric catalog of \citet{2015MNRAS.448.2260G}, based on its blue colors and relatively high proper motion.

We produced an extracted light curve from the processed target pixel file using the {\sc PyKE} software package \citep{2012ascl.soft08004S}, with a fixed aperture of 5~pixels. We removed {\em K2} motion-induced noise with the {\sc kepsff} task \citep{2014PASP..126..948V}. Our final 74.84~d light curve has $108{,}454$ points and a duty cycle of nearly 98.7\%, after iteratively clipping all points falling $>$5$\sigma$ from the mean with the {\sc VARTOOLS} software package \citep{2016A&C....17....1H}. All phases in our light curve are relative to the mid-time of the first exposure: 2457139.6008052~BJD$_{\rm TDB}$.

We display in Figure~\ref{fig:ft} a Fourier transform (FT) of our {\em K2} observations, extending past the Nyquist frequency near 8496\,\muhz, based on our 58.85\,s sampling rate. Our FTs have all been oversampled by a factor of 20; all instrumental harmonics of the long-cadence sampling rate (see \citealt{2010ApJ...713L.160G}) have been fit and removed. We show the confidence threshold of the FT determined by simulating $10{,}000$ light curves, wherein we kept the time sampling but randomly shuffled the fluxes, as outlined in \citet{2015MNRAS.451.1701H}. 99.0\% of these synthetic FTs do not have a peak exceeding 0.92\,ppt (where 1\,ppt $=0.1$\%), which we adopt as our 99\% confidence threshold (shown as a red dotted line in Figure~\ref{fig:ft}). This is close to five times the average amplitude of the entire FT: $5\langle {\rm A}\rangle=0.98$\,ppt.

Notably, two significant peaks in the FT fall within 800\,\muhz\ of the Nyquist frequency; in fact, both peaks mirrored above the Nyquist frequency have more than 5\% higher amplitudes than their sub-Nyquist counterparts. To resolve this Nyquist ambiguity we obtained follow-up, time-series photometry at higher cadence over four nights in 2017~January, from both the 2.1-m Otto Struve telescope at McDonald Observatory in West Texas and the 4.1-m SOAR telescope at Cerro Pach\'{o}n in Chile. All observations were obtained using blue-broadband, red-cutoff filters.

Our 2.1-m McDonald data were taken over three nights with the frame-transfer ProEM camera, using 10\,s exposures taken through a 3mm {\em BG40} filter: on 2017~January~23 (4.8\,hr long, variable $\sim$1.6\arcsec\ seeing, thin clouds), 2017~January~24 (2.4\,hr long, variable $\sim$2.0\arcsec\ seeing, clear skies), and 2017~January~26 (2.0\,hr long, variable $\sim$2.3\arcsec\ seeing, clear skies). In addition, we obtained 2.7\,hr of time-series photometry on 2017~January~26 (stable $\sim$1.8\arcsec\ seeing, clear skies) on the 4.1-m SOAR telescope using the Goodman spectrograph in imaging mode, using 17\,s exposures through an {\em S8612} filter. We began collecting our SOAR light curve 2.6 hr before the McDonald data that night, giving us a two-site duty cycle of more than 16\% over 72\,hr.

Light curves of the ground-based photometry were extracted with circular aperture photometry and corrected to the Solar System barycenter using {\sc WQED} \citep{2013ascl.soft04004T}. An FT of the ground-based data is shown in cyan in Figure~\ref{fig:ft}, including the 99\% confidence threshold of 3.6\,ppt, calculated in the same way as the {\em K2} data. The ground-based data have two significant peaks: one at $5116.53\pm0.30$\,\muhz\ ($4.7\pm0.7$\,ppt) and another at $9177.56\pm0.38$\,\muhz\ ($3.7\pm0.7$\,ppt), confirming that the super-Nyquist signals from the {\em K2} data are in fact those in the star. In both cases, daytime observing gaps have conspired to raise an alias peak to the highest peak in the ground-based dataset (the frequency uncertainties quoted are not appropriate estimates of the actual uncertainties due to the presence of aliases).

We note that the amplitudes of the ground-based data are at least 20\% higher than the {\em K2} amplitudes. This is partly a result of different limb-darkening in our bluer ground-based filters than in the {\em Kepler} bandpass \citep{1995ApJ...438..908R}, as well as flux dilution from a nearby ($<$4.5\arcsec, $\Delta$$K_p$$\sim$2.8\,mag) galaxy. More significantly, we expect phase smearing from the 58.85\,s exposures to suppress the {\em K2} amplitudes around $f_1$ by more than 14\% and $f_2$ by more than 40\%.

\begin{deluxetable}{lrrcr}
\tablecolumns{5}
\tablewidth{0.45\textwidth}
\tablecaption{Frequencies present in \tar\
  \label{tab:freq}}
\tablehead{\colhead{ID} & \colhead{Frequency} & \colhead{Period} & \colhead{Amplitude} & \colhead{Phase}
\\ \colhead{} & \colhead{($\mu$Hz)} & \colhead{(s)} & \colhead{(ppt)} & \colhead{(rad/2$\pi$)} }
\startdata
$f_{1a}$ & 5112.5995(41) & 195.59522 & 3.23 & 0.3312(77) \\ 
$f_{1b}$ & 5250.6035(72) & {\bf 190.45430} & 1.85 & 0.837(13) \\ 
\multicolumn{5}{r}{$f_{1b}-f_{1a}=138.004(11)$ \muhz} \\
$f_{1c}$ & 5389.852(13) & 185.53384 & 1.03 & 0.832(24) \\ 
\multicolumn{5}{r}{$f_{1c}-f_{1b}=139.249(20)$ \muhz} \\
\hline
$f_{2a}$ & 9037.205(17) & 110.65368 & 0.76 & 0.921(32) \\ 
$f_{2b}$ & 9161.6178(94) & {\bf 109.15103} & 1.40 & 0.704(18) \\ 
\multicolumn{5}{r}{$f_{2b}-f_{2a}=124.413(27)$ \muhz} \\
$f_{2c}$ & 9286.287(10) & 107.68566 & 1.37 & 0.136(18) \\ 
\multicolumn{5}{r}{$f_{2c}-f_{2b}=124.669(19)$ \muhz}
\enddata
\end{deluxetable}

Informed by our higher-cadence photometry from McDonald and SOAR, we display in Table~\ref{tab:freq} all six pulsation frequencies that we detect in \tar, marking in bold the $m=0$ components. We include one mode --- $f_{2a}$ --- for which we have slightly relaxed the significance threshold, since it falls where we would expect for a component of a rotationally split multiplet (see Section~\ref{sec:rotation}); our synthetic FTs estimate a 13\% confidence in $f_{2a}$. The values in Table~\ref{tab:freq} have been computed with a simultaneous non-linear least squares fit for the frequency, amplitude, and phase to the {\em K2} data using {\sc PERIOD04} \citep{2005CoAst.146...53L}. The amplitudes have a formal uncertainty of 0.16\,ppt. Our period determination is not in the stellar rest frame, but is at high enough precision that it should be corrected for the gravitational redshift and line-of-sight motion of the white dwarf (e.g., \citealt{2014MNRAS.445L..94D}).

\section{Mass Determination from Spectroscopy}
\label{sec:spectroscopy}

With just two independent pulsation modes at 109.15103\,s and 190.45430\,s, we have only limited asteroseismic constraint on the properties of \tar. Therefore, we obtained low-resolution spectra of as many Balmer lines as possible using the Goodman spectrograph on the 4.1-m SOAR telescope \citep{2004SPIE.5492..331C}. Our setup covers the wavelength range $3600-5200$\,\AA\ with a dispersion of roughly 0.8~\AA~pixel$^{-1}$. We used a 3\arcsec\ slit, so our spectral resolution is seeing limited, roughly 4\,\AA\ in 1.4\arcsec\ seeing.

We obtained spectra of \tar\ over two nights with SOAR. On the night of 2016~February~14 we obtained consecutive $2\times600$\,s exposures in 1.4\arcsec\ seeing at an airmass of 1.8, giving us a signal-to-noise ratio (S/N) of 26 per resolution element in the continuum around 4600\,\AA. Subsequently, we obtained $9\times600$\,s exposures on the night of 2017~January~25 in 2.0\arcsec\ seeing at an airmass of 1.4, yielding S/N $=61$.

We optimally extracted all spectra \citep{1986PASP...98..609H} with the {\sc pamela} software package. We subsequently used {\sc molly} \citep{1989PASP..101.1032M} to wavelength calibrate and apply a heliocentric correction. We flux calibrated the 2016~February spectra with the spectrophotometric standard GD~71 and the 2017~January spectra with Feige~67.

\begin{figure}
\centering{\includegraphics[width=0.995\columnwidth]{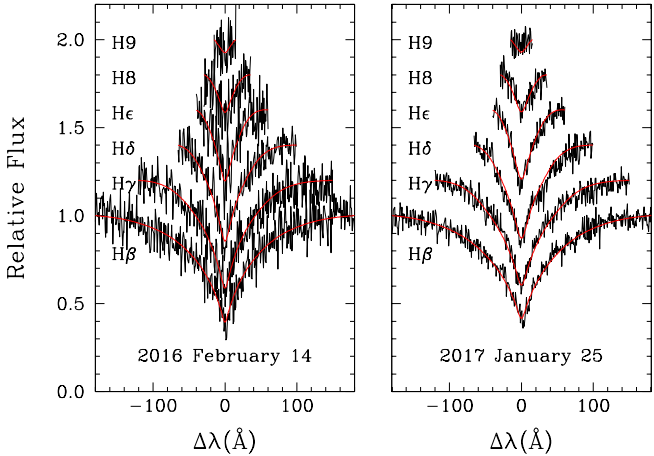}}
\caption{The averaged spectra of $2\times600$\,s exposures the night of 2016~February~14 (left) and of 9$\times$600\,s exposures on 2017~January~25 (right) obtained with the high-throughput Goodman spectrograph on the 4.1-m SOAR telescope. A weighted mean of both fits find that \tar\ is a $13{,}590\pm340$\,K, $0.87\pm0.03$\,\msun\ white dwarf. \label{fig:spec}}
\end{figure}

We fit the six Balmer lines H$\beta$$-$H9 for each epoch of spectroscopy to pure-hydrogen, 1D model atmospheres for white dwarfs that employ the ML2/$\alpha = 0.8$ prescription of the mixing-length theory; the models and fitting procedures are described in \citet{2011ApJ...730..128T} and were convolved to match the resolution set by the seeing. For both epochs we find atmospheric parameters indicating a relatively hot and massive white dwarf: For 2016~February~14 we find 1D parameters of \teff\ $=13{,}010\pm370$\,K, \logg\ $=8.503\pm0.072$, and for 2017~January~25 we find \teff\ $=14{,}020\pm300$\,K, \logg\ $=8.412\pm0.044$. We display the Balmer-line fits in Figure~\ref{fig:spec}.

A weighted mean of both epochs yields 1D atmospheric parameters of \teff\ $=13{,}620\pm340$\,K, \logg\ $=8.437\pm0.052$. We can correct these for the three-dimensional dependence of convection \citep{2013A&A...559A.104T}, which slightly modifies the temperature to $13{,}590$\,K and surface gravity to \logg\ $=8.434$.

The 3D-corrected parameters of \tar\ correspond to a white dwarf mass of $0.87\pm0.03$\,\msun\ using the models of \citet{2013ApJ...779...58R}. If it evolved in isolation, \tar\ would have descended from a roughly $4.0\pm0.5$\,\msun\ B-star progenitor \citep{2016ApJ...818...84C}.

\section{Asteroseismic Measurement of Rotation}
\label{sec:rotation}

\begin{figure}
\centering{\includegraphics[width=0.995\columnwidth]{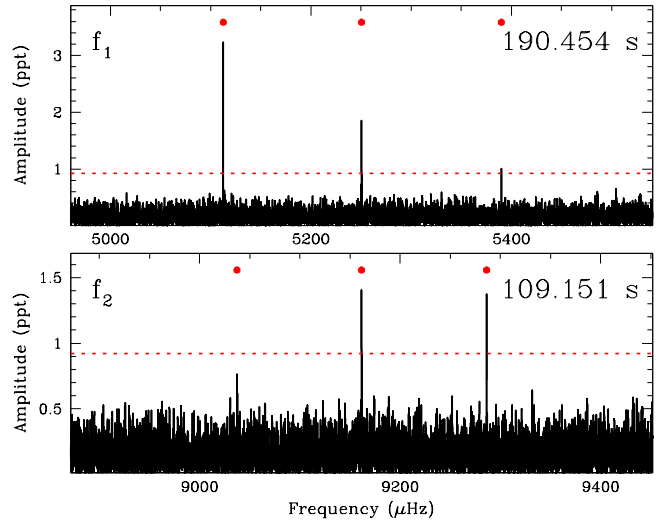}}
\caption{The two main pulsation modes of \tar, centered at $190.45430$\,s and $109.15103$\,s, are most simply interpreted as triplets of $\ell=1$ dipole modes. The modes have weighted mean frequency splittings of $\delta$$f_1=138.626$\,\muhz\ and $\delta$$f_2=124.541$\,\muhz, which correspond to a rotation period of $1.13\pm0.02$\,hr, faster than any pulsating white dwarf known to date. \label{fig:triplets}}
\end{figure}

White dwarfs oscillate in non-radial $g$-modes. In the absence of rotation, pulsations with the same angular degree, $\ell$, and radial overtone, $n$, have the same frequency, independent of the azimuthal order, $m$. Rotation can lift this $m$ degeneracy and decompose a mode into $2\ell+1$ components \citep{1989nos..book.....U}. Due to geometric cancellation of high-$\ell$ modes, we most commonly observe dipole $\ell=1$ modes in white dwarfs, which separate into triplets of $m=-1,0,1$ components.

Thanks to the nearly unblinking, 74-d stare of {\em K2}, we see clearly that the two main modes of \tar\ are each composed of nearly symmetrically split triplets, shown in detail in Figure~\ref{fig:triplets}. We can use the frequency splittings within these modes to estimate the rotation rate of this massive white dwarf. The weighted mean of the frequency splittings for each mode are $\delta$$f_1=138.626\pm0.031$\,\muhz\ and $\delta$$f_2=124.541\pm0.046$\,\muhz.

To first order, we can connect an identified frequency splitting ($\delta$$f$) to the overall stellar rotation ($\Omega$) by the relation $\delta f=m(1-C_{n,\ell})\Omega$, where $C_{n,\ell}$ represents the effect of the Coriolis force on the pulsations as formulated by \citet{1951ApJ...114..373L}. In white dwarfs, $C_{n,\ell}$ is usually close to an asymptotic value of 1/$\ell$($\ell$+1) = 0.5 for $\ell=1$ (e.g., \citealt{1991ApJ...378..326W}). Some modes of lower radial order (especially $n<5$) are strongly affected by abrupt chemical transitions in the layering of the white dwarf, which effectively trap modes to different depths of the star. This trapping causes $C_{n,\ell}$ to deviate below the asymptotic value, and is likely why the mean frequency splittings for $\delta$$f_1$ and $\delta$$f_2$ in \tar\ differ by more than 14\,\muhz\ (10\%).

With just two independent modes, our models are only weakly constrained by asteroseismology. However, guided by our spectroscopically determined effective temperature and overall mass, we have used the evolutionary sequences described in \citet{2013ApJ...779...58R} to explore model-dependent values for $C_{n,\ell}$. We identify $f_1$ at 109.151\,s as an $\ell=1,n=1$ mode and $f_2$ at 190.454\,s as an $\ell=1,n=2$ mode. The $13{,}590$\,K, 0.87\,\msun\ model with a canonically thick hydrogen-layer mass computed in \citet{2013ApJ...779...58R} predicts $C_{1,1}=0.495$ and $C_{2,1}=0.438$ (with mode periods of 98.05\,s and 170.78\,s, respectively). From these model-based $C_{n,\ell}$ values, each triplet {\em independently} yields a rotation period of exactly 1.13\,hr, from both $\delta$$f_1$ {\em and} $\delta$$f_2$.

Our uncertainties on the rotation rate are dominated by the model uncertainties in computing $C_{n,\ell}$, which consistently predict the $\ell=1,n=2$ mode is highly trapped. The most deviant model from \citet{2013ApJ...779...58R} within our spectroscopic uncertainties ($13{,}590$\,K, 0.85\,\msun) predicts $C_{1,1}=0.495$ and $C_{2,1}=0.428$, yielding a rotation period of 1.15\,hr using $f_2$. Therefore, we adopt a rotation period of $1.13\pm0.02$\,hr for \tar.

Finally, the frequency splittings between prograde ($m=0$ to $m=+1$) relative to retrograde ($m=-1$ to $m=0$) components are asymmetric in \tar. That is, the $m=0$ component is not exactly centered between the $m=\pm1$ components. For $f_1$, the observed $m=0$ component is displaced to lower frequency by $0.622\pm0.043$\,\muhz; for $f_2$ the value is $0.128\pm0.073$\,\muhz. The asymmetry is $<$0.5\% and does not significantly affect our inferred rotation rate, but is noteworthy because it likely represents second-order rotation effects, which are expected to be present for such a rapid rotator \citep{1978AcA....28..441C}. A systematic shift of the $m=0$ components can also constrain the presence of a magnetic field too weak to detect from Zeeman splitting of spectroscopy \citep{1989ApJ...336..403J}.

\section{Discussion and Conclusions}
\label{sec:conclusion}

Using {\em K2}, we have discovered a \teff\ $=13{,}590\pm340$\,K, $0.87\pm0.03$\,\msun\ white dwarf with a rotation period of $1.13\pm0.02$\,hr, faster than any known isolated pulsating white dwarf.

\begin{figure}
\centering{\includegraphics[width=0.995\columnwidth]{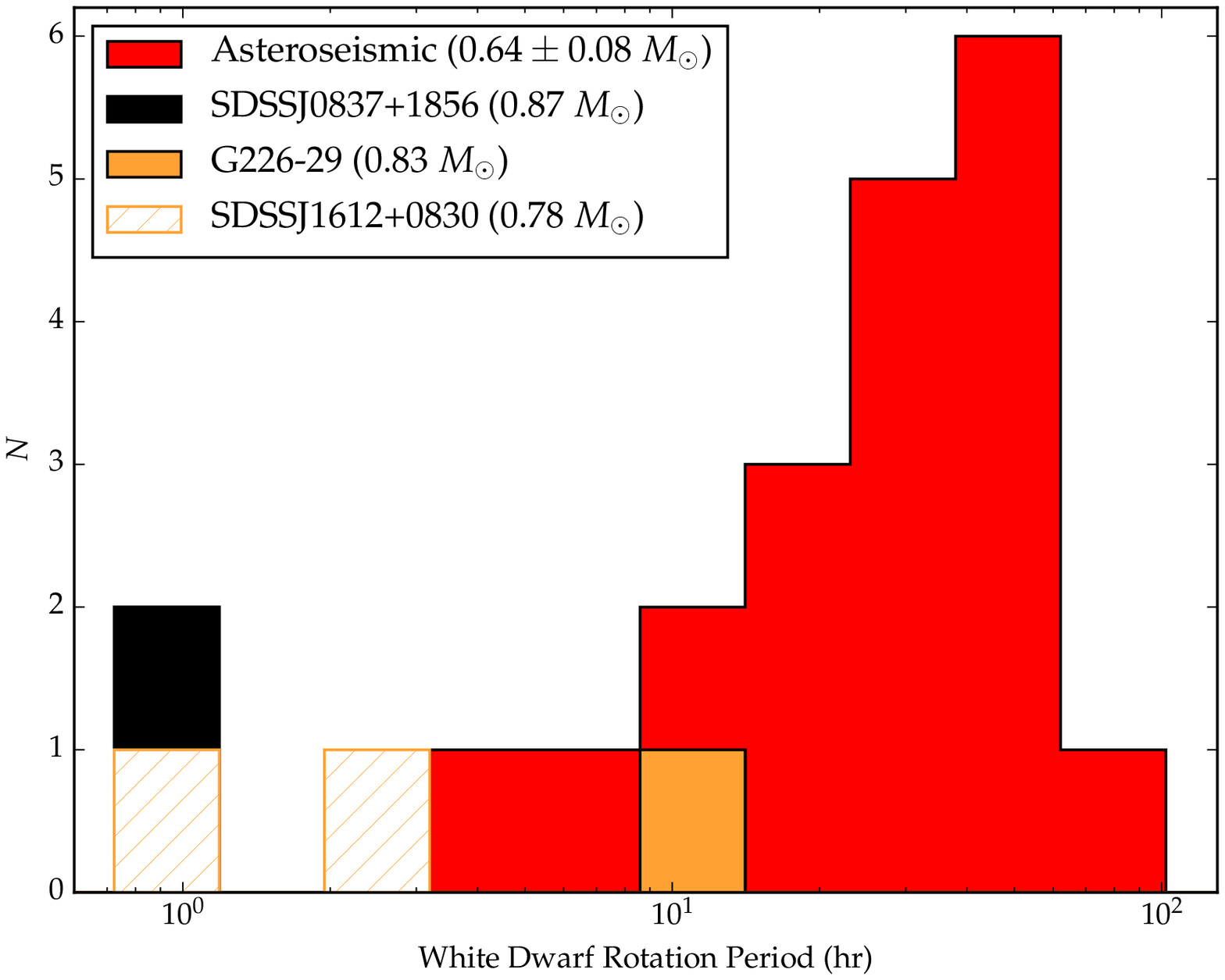}}
\caption{Histogram of rotation rates determined from asteroseismology of pulsating white dwarfs (marked in red), as collected by \citet{2015ASPC..493...65K}. \tar\ (marked in black) is more massive ($0.87\pm0.03$\,\msun) and rotates faster ($1.13\pm0.02$\,hr) than any other pulsating white dwarf known. We mark in orange the only two other pulsating white dwarfs $>$0.75\,\msun\ with rotation estimates. SDSSJ1612+0830 \citep{2013MNRAS.430...50C} is not cleanly identified and could rotate at either 1.0\,hr or 2.0\,hr (see text). \label{fig:rothist}}
\end{figure}

To put this rotation in context, we show in Figure~\ref{fig:rothist} asteroseismically deduced rotation rates of apparently isolated white dwarfs with cleanly identified pulsations from the literature, as compiled by \citet{2015ASPC..493...65K}: GD\,154, HL\,Tau\,76, KUV\,11370+4222, HS\,0507+0434, L19-2, LP\,133-144, GD\,165, R548, G185-32, G226-29, EC14012-1446, KIC\,4552982 \citep{2015ApJ...809...14B}, and KIC\,11911480 (DAVs); PG\,0122+200, PG\,2131+066, NGC\,1501, PG\,1159-035, and RX\,J2117.1+3412 (DOVs); and the DBVs KIC\,8626021 and PG\,0112+104 \citep{2017ApJ...835..277H}. We have excluded the 0.60\,\msun\ white dwarf SDSSJ1136+0409, which rotates at $2.49\pm0.53$\,hr but is currently in a 6.9-hr binary with a detached, nearby dM companion; it underwent binary interaction via a common-envelope event \citep{2015MNRAS.451.1701H}.

All 20 of the white dwarfs with previously published rotation rates likely evolved in isolation, and span the range of rotation periods between roughly $0.4-2.2$\,d. All have spectroscopically deduced atmospheric parameters; the sample has a mean mass of 0.64\,\msun\ and a standard deviation of 0.08\,\msun, including 3D corrections \citep{2013A&A...559A.104T}. This is remarkably similar to the 0.62\,\msun\ mean mass of field white dwarfs \citep{2016MNRAS.461.2100T}, suggesting that the majority of isolated, canonical-mass white dwarfs rotate at $0.4-2.2$\,d.

Figure~\ref{fig:rothist} brings one interesting fact into focus: The three most massive single white dwarfs with rotation measured via asteroseismology are also among the fastest rotators. SDSSJ161218.08+083028.2 (hereafter SDSSJ1612+0830) has significant pulsation periods at 115.17\,s and 117.21\,s \citep{2013MNRAS.430...50C}. These pulsation modes are not cleanly identified, but are too close together to be different radial orders. The simplest explanation is that they are components of a single $\ell=1$ mode split by either 75.6\,\muhz\ or 151.2\,\muhz, indicating a rotation period of roughly 2.0\,hr or 1.0\,hr, respectively; we mark both solutions in Figure~\ref{fig:rothist}. We have refit the SDSS spectrum of SDSSJ1612+0830 using the same models and 3D corrections described in Section~\ref{sec:spectroscopy} and find it has \teff\ $=11{,}800\pm170$\,K, \logg\ $=8.281\pm0.048$, corresponding to a mass of $0.78\pm0.03$\,\msun. Other than \tar, the only star in Figure~\ref{fig:rothist} more massive than SDSSJ1612+0830 is G226-29 ($0.83\pm0.03$\,\msun), which has a single $\ell=1$ mode centered at 109.28\,s with splittings that indicate 8.9-hr rotation \citep{1995ApJ...447..874K}.

It is also possible to measure the rotation of white dwarfs from magnetic spots. There are four spotted white dwarfs with rotation rates shorter than 1\,hr: RE\,J0317$-$853 (12.1\,min, \citealt{2003ApJ...593.1040V}), NLTT\,12758B (22.6\,min, \citealt{2017MNRAS.466.1127K}), SDSSJ152934.98+292801.9 (38.1\,min, \citealt{2015ApJ...814L..31K}), and G99$-$47 (58.2\,min, \citealt{1989LNP...328..329B}). RE\,J0317$-$853 is especially noteworthy since it has a mass inferred from a parallax distance of at least $1.28$\,\msun\ \citep{2010A&A...524A..36K}; its extremely high magnetic field ($>$$200$\,MG) suggests it could be the outcome of a binary merger (e.g., \citealt{2012ApJ...749...25G}).

From population synthesis estimates, roughly $7-23$\% of all apparently single white dwarfs are expected to be the byproducts of mergers \citep{2017arXiv170306893T}. Therefore, it is possible that the white dwarf discovered here, \tar, is not the descendent of single-star evolution. Asteroseismology may rule out this scenario. The 0.877\,\msun\ model from \citet{2012MNRAS.420.1462R}, highlighted in their Figure~6, shows it is difficult to observe an $\ell=1, n=1, m=0$ mode with a pulsation period below 110\,s without the white dwarf having a canonically thick hydrogen layer ($\gtrsim$$10^{-5}$\,$M_{\rm H}/$\mstar). For this reason, we prefer a single-star evolutionary model; however, more asteroseismic analysis and modeling is required to definitively rule out a binary-merger origin.

Finally, we note that \tar\ now supplants HS\,1531+7436 as the hottest known isolated DAV; HS\,1531+7436 has 3D-corrected atmospheric parameters of \teff\ $=13{,}270\pm290$\,K, \logg\ $=8.49\pm0.06$, found using the same model atmospheres \citep{2011ApJ...743..138G}. However, the best-fit effective temperatures for \tar\ from spectra taken on two different nights differ by more than 1000\,K, a $>$2$\sigma$ disagreement (the surface gravities are consistent within the uncertainties). Unfortunately the photometric colors --- ($u$$-$$g$, $g$$-$$r$) $=$ ($0.36\pm0.03$, $-$$0.21\pm0.02$) --- do not strongly prefer one solution over the other \citep{2014ApJ...796..128G}. Given the defining role \tar\ may play in setting the blue edge of the DAV instability strip where pulsations driven by hydrogen partial-ionization finally reach observable amplitudes, it is worth more detailed follow-up spectroscopy to obtain a reliably accurate effective temperature.

\acknowledgments

We acknowledge helpful comments from the anonymous referee, as well as useful discussions with Conny Aerts and Jamie Tayar. 
Support for this work was provided by NASA through Hubble Fellowship grant \#HST-HF2-51357.001-A, awarded by the Space Telescope Science Institute, which is operated by the Association of Universities for Research in Astronomy, Incorporated, under NASA contract NAS5-26555;
NASA {\em K2} Cycle 4 Grant NNX17AE92G;
NASA {\em K2} Cycle 2 Grant NNX16AE54G;
CNPq and PRONEX-FAPERGS/CNPq (Brazil);
NSF grants AST-1413001 and AST-1312983;
the European Research Council under the European Union's Seventh Framework Programme (FP/2007-2013) / ERC Grant Agreement n. 320964 (WDTracer); and
the European Union's Horizon 2020 Research and Innovation Programme / ERC Grant Agreement n. 677706 (WD3D).
Based on observations obtained at the Southern Astrophysical Research (SOAR) telescope, which is a joint project of the Minist\'{e}rio da Ci\^{e}ncia, Tecnologia, e Inova\c{c}\~{a}o da Rep\'{u}blica Federativa do Brasil, the U.S. National Optical Astronomy Observatory, the University of North Carolina at Chapel Hill, and Michigan State University, as well as data taken at The McDonald Observatory of The University of Texas at Austin.

{\it Facilities:} {\em K2}, SOAR (Goodman), Otto Struve (ProEM)


\end{document}